\documentclass[12pt]{iopart}

\usepackage{graphicx}
\usepackage{color}     
\usepackage{ulem}
\usepackage{hyperref}

\newcommand{\CM}[1]{\textcolor{black}{{#1}}}

\newif\ifCMbool
\CMboolfalse

\ifCMbool
\newcommand{\CMs}[1]{\textcolor{red}{{\sout{#1}}}}
\else
\newcommand{\CMs}[1]{\textcolor{red}{{ }}}
\fi

\begin{document}

\title[Phase transition of IrTe$_2$ measured by XPD]{Break of symmetry at the surface of IrTe$_2$ upon phase transition measured by X-ray photoelectron diffraction}

\author{Maxime~Rumo$^{1,}$*, Aki~Pulkkinen$^{1,2,}$, KeYuan~Ma$^{3,}$, Fabian~O.~von~Rohr$^{3,}$, Matthias~Muntwiler$^{4,}$ and Claude~Monney$^{1,}$*}

\address{$^{1}$ \quad D{\'e}partement de Physique and Fribourg Center for Nanomaterials, Universit{\'e} de Fribourg, CH-1700 Fribourg, Switzerland}
\address{$^{2}$ \quad School of Engineering Science, LUT University, FI-53850 Lappeenranta, Finland}
\address{$^{3}$ \quad Department of Chemistry, University of Zurich, CH-8057 Zurich, Switzerland}
\address{$^{4}$ \quad Paul Scherrer Institute, CH-5232 Villigen PSI, Switzerland}

\ead{maxime.rumo@unifr.ch \& claude.monney@unifr.ch}

\vspace{10pt}
\begin{indented}
\item[]August 2021
\end{indented}

\begin{abstract}
IrTe$_2$ undergoes a series of charge-ordered phase transitions below room temperature that are characterized by the formation of stripes of Ir dimers of different periodicities. Full hemispherical X-ray photoelectron diffraction (XPD) experiments have been performed to investigate the atomic position changes undergone near the surface of $1T-$IrTe$_2$ in the first-order phase transition, from the $(1\times1)$ phase to the $(5\times1)$ phase. Comparison between experiment and simulation allows us to identify the consequence of the dimerization on the Ir atoms local environment. We report that XPD permits to unveil the break of symmetry of IrTe$_2$ trigonal to a monoclonic unit cell and confirm the occurence of the $(5\times1)$ reconstruction within the first few layers below the surface with a staircase-like stacking of dimers.
\end{abstract}

\vspace{2pc}
\noindent{\it Keywords}: Phase Transition, X-ray Photoelectron Diffraction, Local Real Space

\submitto{\JPCM}

\maketitle
 
\ioptwocol

\section{Introduction}

Complete determination of atomic positions is necessary to understand the behavior of material surfaces. From the numerous techniques available to obtain information about crystal structures, X-ray photoelectron diffraction (XPD) has proven to be powerful given its chemical sensitivity and its ability to measure atomic displacements in the subangstrom range, in real space~\cite{Kono_1980,Osterwalder_1987,Osterwalder_Book_2003,DespontSurfScience,DespontEUPhysJB}. Nevertheless, without theoretical simulations it is difficult to understand the different patterns emerging in an XPD result and to correlate them to the real atomic positions. The comparison of simulations and measurements is efficient, as in the present case, because the single-scattering approach considerably accentuates the so-called emitter-scatterer "forward focusing" effect. It allows the interpretation of finer patterns, which are essential for the determination of the changes undergone during phase transitions. 
\\

\begin{figure*}[h!]
\centering
\includegraphics[width=1.65\columnwidth]{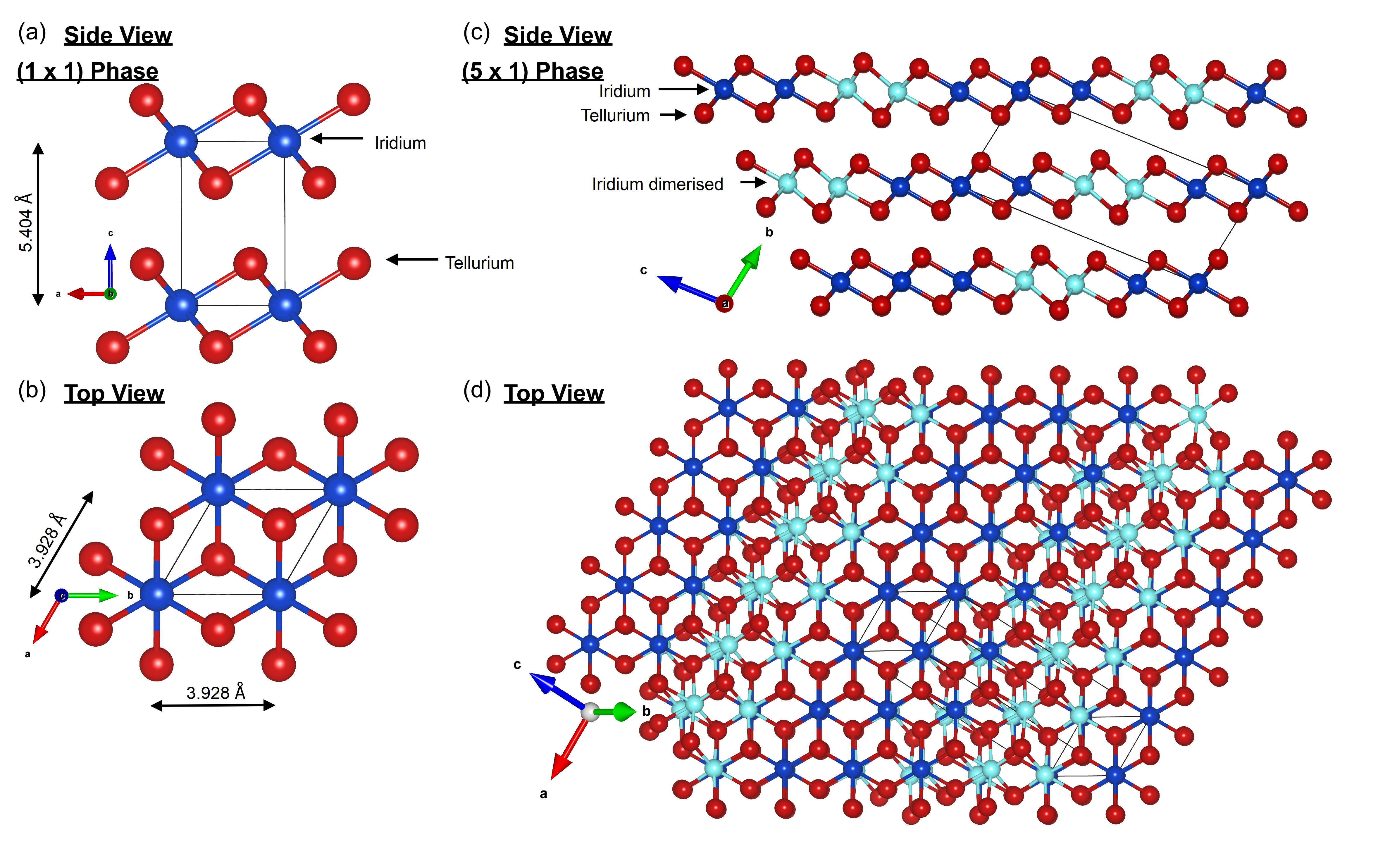}
\caption{\label{Figure Structure} 
(Color online)~(a) Unit cell of $1T-$IrTe$_2$ in the room temperature phase, top view and the side view (b). Ir atom (blue) planes are sandwiched between the Te atoms (red). (c) Unit cell of IrTe$_2$ in the $(5\times1)$ phase, side view and top (d). Dimerised Ir atoms are depicted in light blue.} 
\end{figure*}

Transition metal dichalcogenides have generated a substantial interest for a long time due to their quasi two-dimensional character, interesting electronic properties and various phase transitions. Among them, $1T-$IrTe$_2$ is particularly attractive because of its sandwich-like basic structure composed of hexagonal planes of Ir between two planes of Te giving a quasi two-dimensional structure at room temperature (RT), see Figure~\ref{Figure Structure}~(a). IrTe$_2$ undergoes several structural first-order phase transitions below RT. The system goes from a trigonal unit cell of CdI$_2$-type ($P\overline{3}m1$) to a monoclinic unit cell ($P\overline{1}$) accompanied by a sudden jump in resistivity and magnetic susceptibility at T$_{c_1} = 278$~K~\cite{KoNatCom,FangScienRep,JobicZeit,MatsumotoJLTP,ToriyamaJapanLetters,KoleySSCom,LiSciRep,ParisXAS,RumoPRB}. In this first charge ordered phase, one-dimensional stripes of Ir dimers~\cite{HsuPRL,PascutPRL} appear due to a large decrease of their bond length and lead to a bulk $(5\times1\times5)$ superstructure~\cite{PascutPRL,PascutPRB,KoNatCom,LiNatCom,ToriyamaJapanLetters,OhPRL,TakuboPRB}, see Figure~\ref{Figure Structure}~(c-d). Although the changes in the in-plane bonding suggest a multi-center bond as a more complete description~\cite{SalehEntropy} for brevity we will continue to call them “dimers” throughout the text. A second phase transition occurs at T$_{c_2} = 180$~K, characterized by a bulk $(8\times1\times8)$ superstructure. This has stimulated numerous scanning tunneling microscopy (STM)~\cite{MauererPRB,HsuPRL,KoNatCom,LiNatCom,DaiPRB} and angle-resolved photoemission spectroscopy (ARPES) studies~\cite{RumoPRB,LeeIOP,OotsukiJapanLettersES,OotsukiJournPhys,KoNatCom,NicholsonComMater2021}, which revealed additional periodicities and a surface periodicity $(6\times1)$ appearing after a third phase transition at T$_{c_3} = 165$~K. These surface-sensitive probes disclosed a complex evolution of the electronic structure of IrTe$_2$ through its phase transitions that calls for state-of-the-art ab-initio calculations for a better understanding. This requires the determination of the atomic structure of IrTe$_2$ in the different reconstructed phases up to a few atomic layers below the surface, to be comparable to the probing depth of typical ARPES measurements. So far, only bulk-sensitive X-ray diffraction provided such structural information while STM probes mainly the Te layer. XPD is an excellent technique to fill in this gap, since x-ray photoelectrons are emitted within their inelastic mean free path, generally a few nanometres.
\\

Here we investigate the $1T-$IrTe$_2$ surface first-order phase transition, from the $(1\times1)$ phase at room temperature to the $(5\times1)$ phase at $250$~K, using XPD measurements in comparison to simulated diffractograms. We identify the different changes caused by the structure reconstruction and by extension the dimerization of Iridium and Tellurium atoms. This way, we qualitatively confirm the occurrence of the $(5\times1)$ reconstruction within the first few layers below the surface with a staircase-like stacking of dimers.

\section{Methods}


Single crystals of IrTe$_2$ were grown using the self-flux method~\cite{FangScienRep,JobicZeit}. They were characterized by magnetic susceptibility and resistivity measurements, which confirm that T$_{c_1}$ $= 278$~K and T$_{c_2}$ = $180$~K~\cite{RumoPRB}. Samples were cleaved at RT in vacuum at a pressure of about $10^{-8}$~mbar. During the XPD measurements, the base pressure was better than $5\times10^{-10}$~mbar. The study was done at the PEARL beamline of the Swiss Light Source facility in the Paul Scherrer Institute. The surface quality was checked by X-ray photoemission spectroscopy (XPS). The XPD measurements were all taken on a photoemission station designed as an ARPES facility composed of Carving 2.0 six-axis manipulator and a Scienta EW4000 hemispherical analyser with two-dimensional detection with a $500$~eV photon energy light source. The processing of the angle-scanned XPD data and the normalization procedure is detailed in the PEARL station description~\cite{PEARL}. Note that experimental diffractograms are acquired by simultaneously collecting electrons from $58.9$ to $63.3$~eV. Data has been collected from $0^{\circ}$ up to $90^{\circ}$ polar angle $\theta$ and from $0^{\circ}$ up to $360^{\circ}$ in steps of $40^{\circ}$ azimuth angle $\phi$.

\section{Results \& Discussions}

\subsection{Details on the technique}

By focusing on a particular core level in XPS, it is possible to choose the specific emission intensity of a given atom. The selected outgoing photoemitted electrons have a strong anisotropic distribution of angular intensity related to the local geometry around the target atom. The analysis of the obtained XPD patterns is simplified for electron kinetic energies above about $500$~eV. Above this energy the strong anisotropy in the individual electron-atom scattering leads to a forward focusing of electron flux along directions pointing from the photoemitter to the scatterer~\cite{Fasel_SurfSci_1995}. The so-called forward focusing effect consists of a strong increase in intensity along the emitter-scatterer direction and more generally along densely packed atomic planes (giving so-called Kikuchi bands) and rows of atoms (corresponding to low-index crystallographic directions)~\cite{DespontEUPhysJB,FedchenkoIOP}.

\subsection{Simulated stereographic projection diffractogram}
\label{Simulated stereographic projection diffractogram}

\begin{figure}[t]
\centering
\includegraphics[width=1\columnwidth]{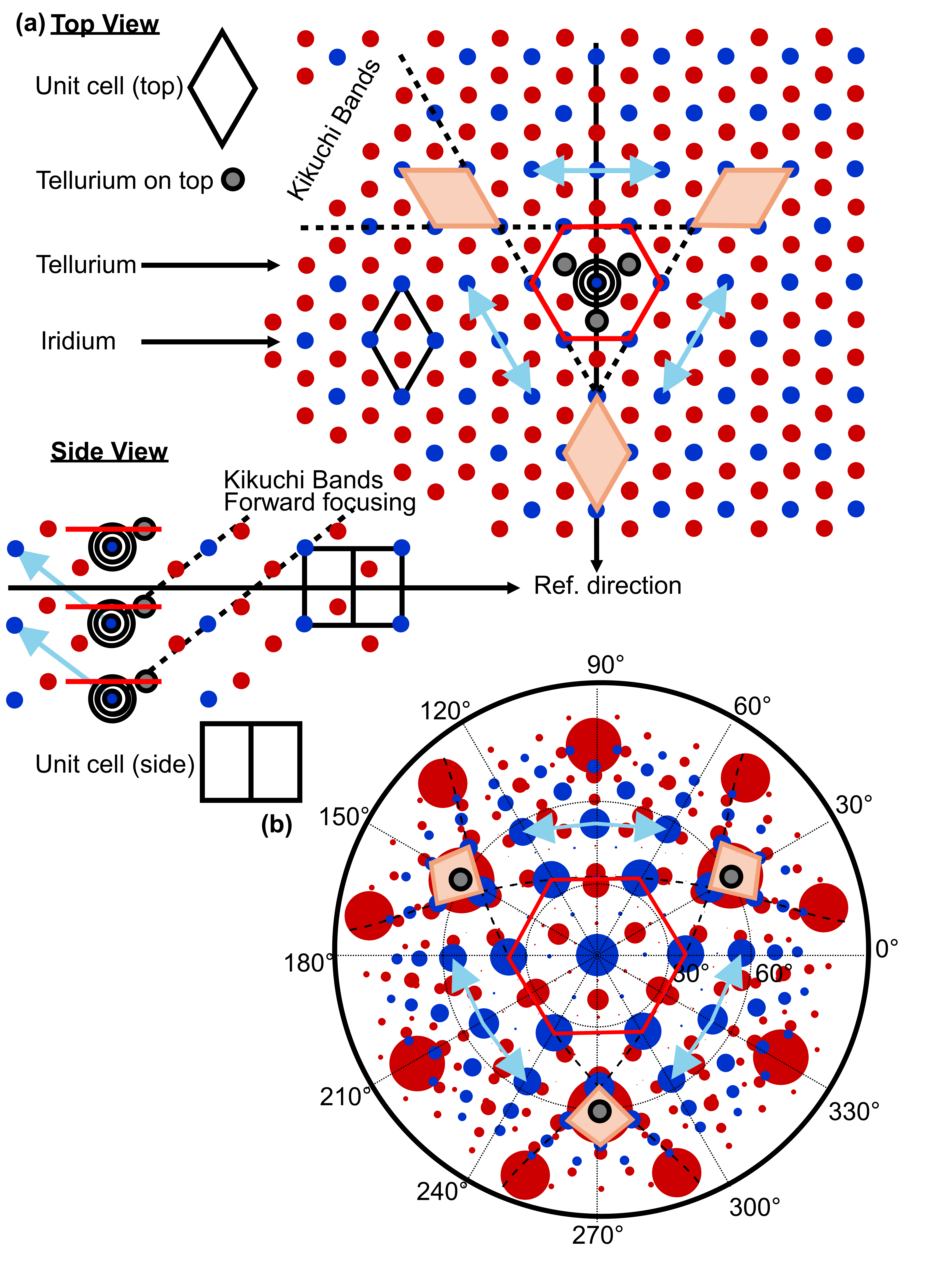}
\caption{\label{Figure Stereographic Proj} 
(Color online)~(a) In plane cut of a hexagonal IrTe$_2$ structure of the $(1\times1)$ phase with the specific patterns described in the text and a side view. (b) Simulated stereographic projection diffractogram for a Ir atom emitter of the $1T-$IrTe$_2$ structure in the $(1\times1)$ phase. The (scatterer) atoms are represented by red bullets for Te and blue for Ir.} 
\end{figure}

In order to anticipate the main intensity peaks in the XPD measurements, we describe below specific patterns that can be observed from a diffractogram emitted from an Iridium atom, later called Ir emitter. Figure~\ref{Figure Stereographic Proj}~(a) shows in-plane and out of plane cuts of the $1T-$IrTe$_2$ crystal structure, where Ir atoms are represented with blue markers and Te atoms with red markers. The Te atoms closest to the Ir emitter, in grey, will lead to major intensity peaks, with the forward-focusing effect, in the XPD diffractograms. There are 6 Ir atoms near the emitter, highlighted with a red hexagon, which will result in high intensities in the XPD diffractograms. In addition, these Ir atoms are also located on the Kikuchi bands $\{101\}$, the black dashed lines. The crossings of the Kikuchi bands, on a batch of Ir atoms, reflect the orientation of the unit cell as well as 3-fold symmetry and are therefore sensitive to modifications of the structure during phase transitions. This crossing is highlighted by an orange quadrilateral. The next closest Ir atoms, highlighted with light blue arrows, will also cause distinctive intensities in the XPD diffractograms. 
\\

To simulate, in a simple way, the diffraction patterns expected from the structure above, we adopt the following strategy. We construct a cluster of atoms centred on an Ir atom, chosen to be an emitter. For the $(1\times1)$ phase, the cluster is developed for the $1T$ crystallographic structure (space group $P\overline{3}m1$) with the lattice parameters $a = b = 3.93$~\AA$ $ and $c = 5.40$~\AA$ $ and a trigonal unit cell, as presented in Figure~\ref{Figure Structure}~(a). The Ir emitters are distributed, one per sandwich, down to $13$~\AA$ $ below the surface which corresponds to three layers, obtained from the electrons inelastic mean free path (IMFP)~\cite{BookMPSeah}. The in-plane size of the cluster consists of $13\times13$ unit cells. For the $(5\times1)$ phase, we take a cluster size similar to the $(1\times1)$ phase cluster (in terms of number of atoms as the unit cell in the $(5\times1)$ phase is very large), with the lattice parameters from the literature $a^{\star} = 3.95$~\AA, $b^{\star} = 6.65$~\AA$ $ and $c^{\star} = 14.45$~\AA$ $ in a monoclinic unit cell ($P\overline{1}$)~\cite{JobicZeit,PascutPRB}. Due to the symmetry break induced by the phase transition (the three fold symmetry switching to an one fold symmetry), there are then 5 Ir atoms per unit cell. According to the Wyckoff classification system, we have one Ir atom (monomer) at $(1a)$ site, two Ir atoms (monomer) at $(2i)$ sites and two Ir (dimer) at $(2i)$ sites. We attribute these different sites to 5 inequivalent Ir atoms, based on a local view. We then discriminate these inequivalent atoms in the calculations to be able to construct the different diffractograms representing the structure from the perspective of a monomer or dimer emitter.
\\

For a cluster centred on an Ir emitter, as described above, we build a simulated diffractogram as follows. For all scattering atoms above this emitter, we calculate the relative position in polar coordinates. We display in Figure~\ref{Figure Stereographic Proj}~(b) in stereographic projection these relative positions in order to simulate the XPD diffractograms. For this purpose, we calculate the relative distance of these scattering atoms $k$ to the emitter $L_k$, while noting the atomic mass $Z$, in addition to the distance of the emitters $i$ to the surface $d_i$, in order to determine their respective sizes. The size of the scattering intensities $D_{k,i}$ in the stimulated diffractograms is then established from the following formula, knowing that for neighbouring atoms of the same type (angular distances less than 0.01) we sum their respective sizes,

\begin{center}
\begin{equation*}
D_{k,i} = Z e^{\frac{-L_k - d_i}{\lambda}},
\end{equation*}
\end{center}

where $\lambda$ is a proportional factor related to the IMFP ($\lambda = 13$~\AA). The scattering contributions of all three layers and all inequivalent emitters are finally summed to generate simulated diffractograms. All the features highlighted in Figure~\ref{Figure Stereographic Proj}~(a) are also present in the simulated diffractogram in Figure~\ref{Figure Stereographic Proj}~(b) with the same color code. An additional effect is the refraction that occurs as the electron wave passes from the solid to the vacuum. We therefore introduce a correction $\theta^{cor}$ on the polar angle $\theta$ in proportion to the kinetic energy ($E_{kin}$) and the inner potential ($V_0=$ $13$~eV~\cite{KoNatCom,LeeIOP}) which is defined as

\begin{center}
\begin{equation*}
\label{XPD polar correction}
\theta^{cor} = \arcsin \left( \sin(\theta)\sqrt{\frac{E_{kin}}{E_{kin}-V_0}} \right).
\end{equation*}
\end{center}

The contributions of each type of scatterer \lbrack Ir (monomer), Te and Ir (dimer)\rbrack ~have been differentiated by color in the simulated diffractograms. This enables the precise identification of the origin of the intensities in the XPD measurements, but reduces the agreement with the XPD measurements. Indeed, the bullets therefore appear smaller and denser in the simulated diffractograms in the ($5\times1$) phase, as $2/5$ of the Ir atoms are dimerised with a strongly modified structure compared to the RT structure, see Figure~\ref{Figure Structure}.

\subsection{X-ray photoemission spectroscopy}

\begin{figure}[h!]
\centering
\includegraphics[width=1\columnwidth]{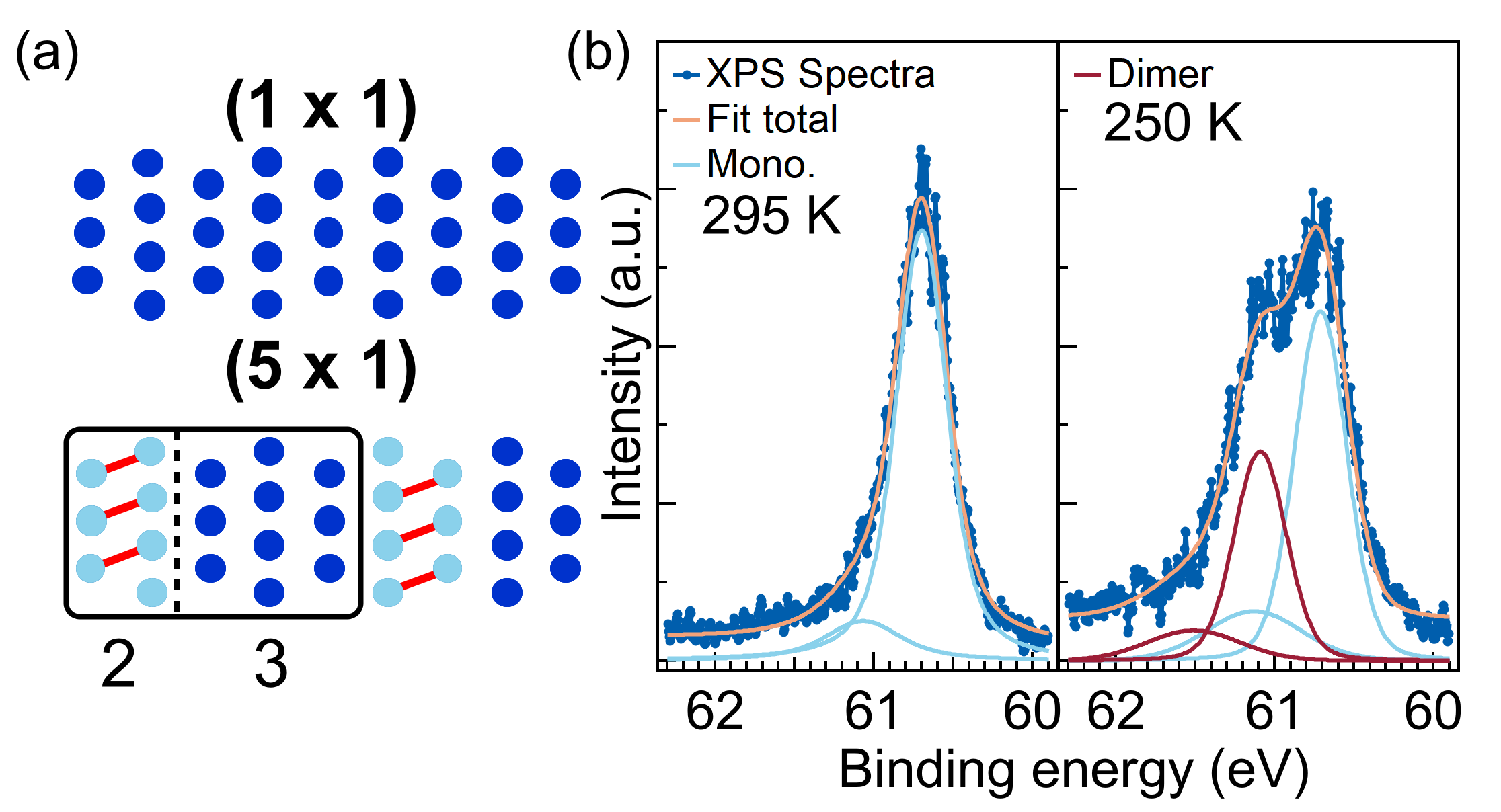}
\caption{\label{Figure Fit - XPS} 
(Color online)~(a) Schematic representation of the atomic structure in the Ir planes for the $(1\times1)$ and the $(5\times1)$ phase. (b) XPS spectra of Ir~$4f_{7/2}$ core levels measured with a $h\nu = 500$~eV photon energy at $295$~K and $250$~K along with the fit components, both taken at $25^{\circ}$ in polar angle $\theta$ and $150^{\circ}$ in azimuthal angle $\phi$.} 
\end{figure}

In Figure~\ref{Figure Fit - XPS}~(a), we recall schematically the structure of the basic building blocks for the Ir plane in two different phases of IrTe$_2$. The ($1\times1$) phase is composed only of equivalent Ir atoms represented with blue bullets [as seen also in Figure~\ref{Figure Structure}~(a-b)], leading to a single Ir $4f_{7/2}$ core level at $60.6$~eV. In the ($5\times1$) phase, 5 atoms split in 2 dimerized Ir atoms represented with light blue bullets (one dimer) and 3 undimerized atoms [Figure~\ref{Figure Fit - XPS}~(a) \& Figure~\ref{Figure Structure}~(c-d)]. This leads to a splitting of the Ir $4f_{7/2}$ core level into a contribution due to the dimerized Ir atoms~\cite{KoNatCom,QianIOP,RumoPRB,NicholsonComMater2021}, at higher binding energy, and a contribution due to the 3 undimerized Ir atoms. These different contributions in XPS are used to acquire XPD diffractograms specific to the local environment of the monomer (undimerised) and dimer Ir atoms. We present the corresponding XPS spectra and detailed fit of the Ir $4f$ core levels in Figure~\ref{Figure Fit - XPS}~(b) with a zoom on the Ir $4f_{7/2}$ core levels measured at different temperature upon cooling. A clear splitting occurs below T$_{c_1}$ between the peak at $60.6$~eV binding energy, which is attributed to the monomer states, and a new peak appearing at $61.2$~eV binding energy corresponding to the dimer states. The intensity ratio measured by XPS has been interpreted as a measure of the density of dimers in the different phases observed in IrTe$_2$~\cite{KoNatCom,QianIOP,RumoPRB}. Below T$_{c_1}$, in the $(5\times1)$ phase, the dimerized Ir atom ratio is $0.4$ and this is also in good agreement with the relative peak intensities in our XPS data at normal emission.
\\

Figure~\ref{Figure Fit - XPS}~(b) displays also the fit components used in the data processing. The panels show an exemplary spectrum of the full XPD data at two different temperatures. In the first panel, the raw data (in blue) at room temperature (RT) in the $(1\times1)$ phase are fitted with two Voigt functions, one is attributed to the main core level and the second to a shake-up satellite\footnote{An asymmetric line shape was already observed in metallic IrO$_2$ and explained in the photoemission process~\cite{PfeiferXPS}. By analogy, we use a similar fit model.} (in light blue), composing the total fit (in orange). The Voigt functions have a fixed position, at $60.6$~eV and $61.0$~eV respectively, and their relative amplitudes are fixed. The second panel displays an exemplary spectrum of the full XPD data taken at $250$~K in the $(5\times1)$ phase (in blue) along with the fit components. Since IrTe$_2$ has undergone a phase transition expressed in a splitting of the Ir $4f_{7/2}$ core level, we then chose to use four Voigt functions with fixed position and with their relative amplitudes fixed two by two. Thanks to this fitting procedure, out of a full XPD data set we obtain two XPD diffractograms of IrTe$_2$ in the $(5\times1)$ phase. An XPD diffractogram for the emitter in the monomer state and the second one for the emitter in the dimer state are then generated. In order to simplify the language, we will use the term monomer emitter for an Ir emitter atom in monomer state and dimer emitter for an Ir emitter atom in the dimer state. 

\subsection{X-ray photoelectron diffraction}

\begin{figure*}[h!]
\centering
\includegraphics[width=1.8\columnwidth]{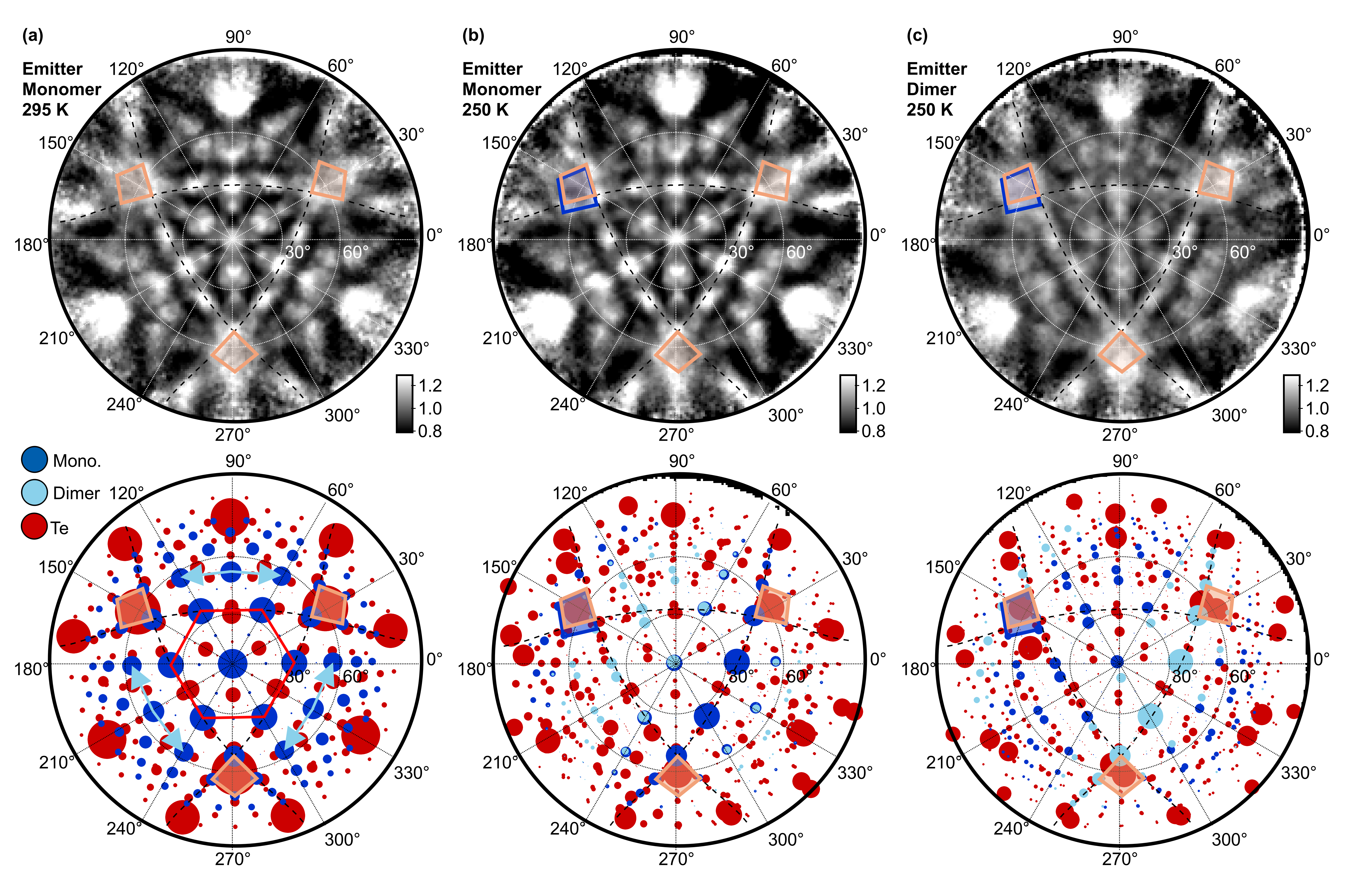}
\caption{\label{Figure XPD measurement} 
(Color online)~Experimental XPD diffractograms presented as two-dimensional gray scale intensity maps with their respective simulated diffractograms in stereographic projection. (a) Diffractograms for a monomer emitter in the ($1\times1)$ phase at $295$~K, (b) for a monomer emitter in the ($5\times1)$ phase at $250$~K and (c) for a dimer emitter in the ($5\times1)$ phase at $250$~K. The scatterer atoms are represented with dark blue bullet for monomer Ir atoms, light blue for dimer Ir atoms and red for Te atoms.} 
\end{figure*}

Figure~\ref{Figure XPD measurement}~(a-c) displays the XPD diffractograms (top row) and their respective simulated diffractograms (bottom row) for an Ir $4f_{7/2}$ core level emitter in the $(1\times1)$ phase at $295$~K from a monomer emitter, in the $(5\times1)$ phase from a monomer emitter and in the $(5\times1)$ phase from a dimer emitter at $250$~K, acquired using $500$~eV photon energy. These measurements and simulations probe the local environment in real space for both monomer and dimer emitters. Normal emission intensity is at the center and grazing angle emission is at the edge of the diffractogram. The XPD measurements present the high intensities in white and the low intensities in black. The simulated diffractograms are constructed according to the procedure detailed in section~\ref{Simulated stereographic projection diffractogram}, where the Te scatterer atoms are represented with red bullets, the Ir monomer scatterer atoms with dark blue bullets and the Ir dimer scatterer atoms with light blue bullets.
\\

The diffractograms in the $(1\times1)$ phase show the typical 3-fold symmetry of the IrTe$_2$ space group. In the experimental data from Figure~\ref{Figure XPD measurement}~(a), the forward focusing peaks, present since electrons have a kinetic energy of $500$~eV, are dominant and define intensity patterns that reflect the emitter local environment. These peaks are located along low-index directions, as well as on the Kikuchi bands connecting these directions. The patterns at the crossing of the Kikuchi bands are highlighted in orange. The particular patterns anticipated in section~\ref{Simulated stereographic projection diffractogram} can be recognized in the XPD diffractograms, see Figure~\ref{Figure Stereographic Proj}, from the signatures of the closest Ir atoms close to $30^{\circ}$ in polar angle $\theta$ (red hexagon) to the next closest Ir atoms below $60^{\circ}$ in polar angle $\theta$ at $90^{\circ}$, $210^{\circ}$ and $330^{\circ}$ in azimuth angle $\phi$ (light blue) and in particular the crossings of the Kikuchi bands, as already mentioned. The simulated diffractogram in Figure~\ref{Figure XPD measurement}~(a) contains only fingerprints originating from Ir monomer and Te atoms, since the crystal is in the $(1\times1)$ phase. Globally, the simulation is comparable to the experimental measurement.
\\

Now we focus on the main pattern modifications in XPD diffractograms for a monomer emitter in the $(5\times1)$ phase, see Figure~\ref{Figure XPD measurement}~(b). Although the experimental XPD diffractogram for a monomer emitter in the $(5\times1)$ phase have similarities with the experimental XPD diffractogram for a monomer emitter in the $(1\times1)$ phase, we observe a symmetry break from the trigonal to the monoclinic unit cell that occurs in particular at the intersection of the Kikuchi bands at about $150^{\circ}$ in azimuth angle $\phi$. This change is pointed out with a blue quadrilateral. The Kikuchi bands crossing is shifted by a few degrees in azimuth angle $\phi$ towards $180^{\circ}$. In contrast, the Ir atoms plane on the Kikuchi band at the opposite side of this crossing is not much affected by the phase transition. The respective local environments of the different inequivalent Ir atoms\footnote{The specific simulated diffractogram for each inequivalent emitter are shown in the supplementary.} are affected by the phase transition which explains the changes in the simulation. This is the result of structural modifications by the emergence of dimerised Ir atoms in the structure as shown in the top view of Figure~\ref{Figure Structure}~(d). For example, the dimerised Ir atoms in the layer above the inequivalent monomer emitters are located more on the left side rather than the other and this results in a higher concentration of dimer signatures in upper left part of the simulated diffractogram. Moreover, in the simulated diffractogram, the stripes of dimerised Ir atoms are arranged along the one-dimensional axis from $60^{\circ}$ to $240^{\circ}$ in azimuth angle $\phi$, in the same direction as in Figure~\ref{Figure Structure}~(d). These split signatures, in the simulated diffractogram, are expressed as shifts in intensity, in the experimental XPD diffractogram, and are in good agreement. This suggests then that the experimental measurement shows a preferential direction of dimerisation.
\\

The experimental XPD diffractogram for dimer emitters, in the $(5\times1)$ phase in Figure~\ref{Figure XPD measurement}~(c), displays broader intensities. Although some of the patterns strongly remind those of the monomer XPD diffractograms  such as the Kikuchi bands and the highlighted quadrilaterals, they are more difficult to identify. The simulated diffractogram, in Figure~\ref{Figure XPD measurement}~(c), shows the same perturbation of the Kikuchi band crossing as well as a stronger splitting of the Ir monomer scatterer, Ir dimer scatterer and Te scatterer signatures as in the simulated diffractogram for the monomer emitter in Figure~\ref{Figure XPD measurement}~(b). The signatures of Te atoms at $90^{\circ}$, $210^{\circ}$ and $330^{\circ}$ in azimuth angle $\phi$ near $77^{\circ}$ in polar angle $\theta$ are significantly perturbed. The structure of the Figure~\ref{Figure Structure}~(c) provides a better understanding of the local environment of the dimer emitters. The dimerised Ir atom local environment changes are then enhanced in the simulation as well as the diffraction measurement since the dimerised Ir atom moved by $0.8~$\AA~\cite{PascutPRL} from its original RT position, with staircase-like stacking. Dimerised Ir atoms therefore have an apparent disordered local environment, see Figure~\ref{Figure Structure}~(d). The resulting patterns in the experimental and simulated diffractograms, in the $(5\times1)$ phase, for a dimer emitter, are then broader than the other experimental and simulated diffractograms. 
\\

\begin{figure}[h!]
\centering
\includegraphics[width=1\columnwidth]{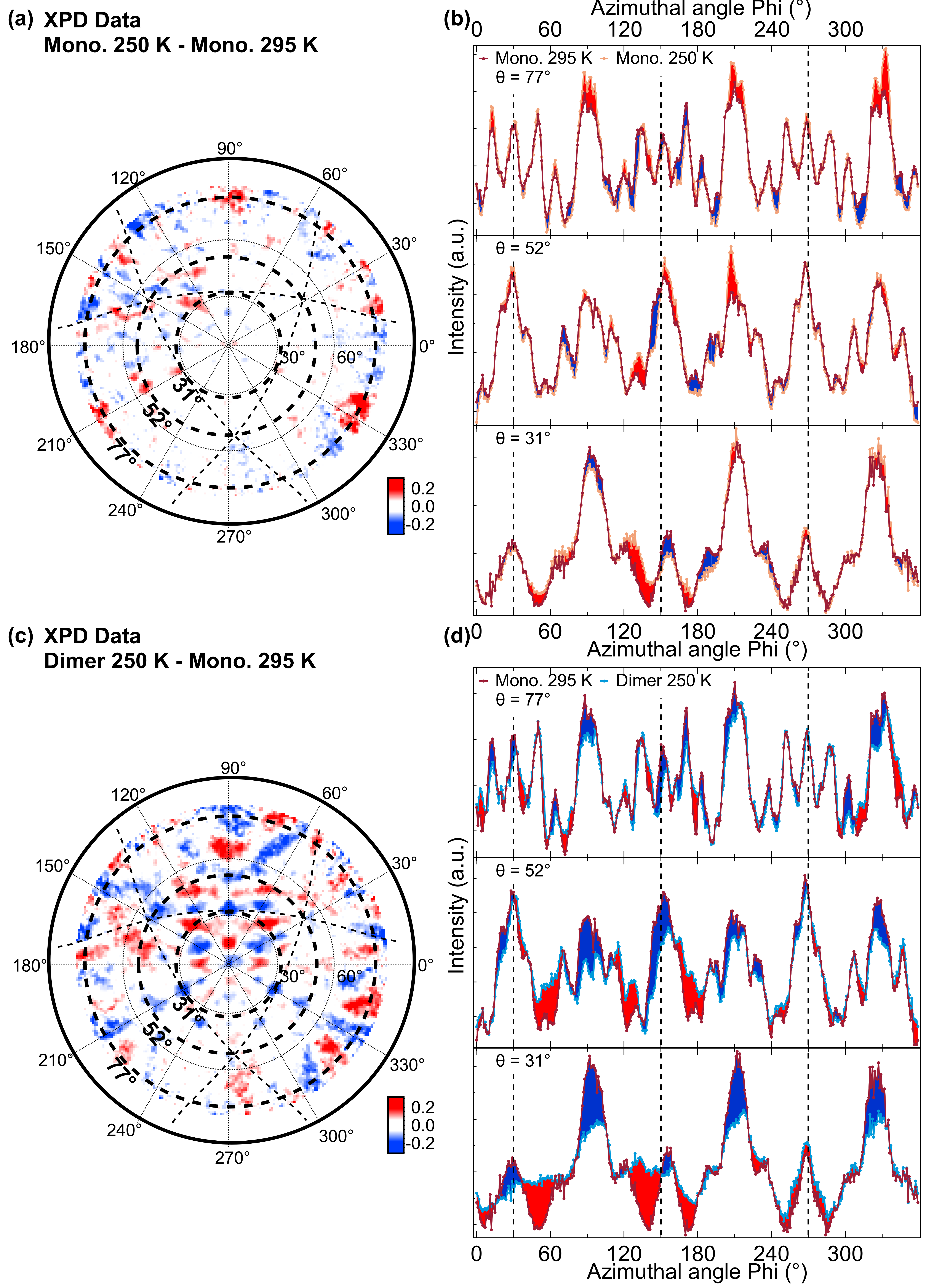}
\caption{\label{Figure Exp Azimuth} 
(Color online)~(a) Diffractogram constructed as the experimental XPD diffractogram for a monomer emitter in the $(5\times1)$ phase minus the experimental XPD diffractogram for a monomer emitter in the $(1\times1)$ phase. (b) Detailed azimuthal cuts of the (a) diffractogram as a function of polar angle $\theta$. (c) Diffractogram constructed as the experimental XPD diffractogram for a dimer emitter in the $(5\times1)$ phase minus the experimental XPD diffractogram for a monomer emitter in the $(1\times1)$ phase. (d) Detailed azimuthal cuts of the (c) diffractograms as a function of polar angle $\theta$. The gain and losses are respectively displayed in red and blue in the diffractograms.} 
\end{figure}

To emphasise the intensity changes undergone across the first phase transition, Figure~\ref{Figure Exp Azimuth}~(a) displays a diffractogram\footnote{We cut the diffractogram above $81^{\circ}$ in polar angle $\theta$ to eliminate noisy data near the grazing angles.} constructed with the experimental XPD diffractogram for an monomer emitter in the $(5\times1)$ phase minus the experimental XPD diffractogram for a monomer emitter in the $(1\times1)$ phase. The azimuthal cuts at three different polar angles $\theta$ shown in Figure~\ref{Figure Exp Azimuth}~(b) display the gains (in red) and losses (in blue) of intensities caused by the phase transition. We see that the main variation is at the crossing of the Kikuchi bands near $150^{\circ}$ in azimuth angle $\phi$ near $52^{\circ}$ in polar angle $\theta$, as observed in Figure~\ref{Figure XPD measurement}. Intensity from below $150^{\circ}$ in azimuth angle $\phi$ is transferred to higher azimuthal angle. This is due again to the breaking of the three-fold symmetry across the phase transition. A significant gain is observed at $90^{\circ}$, $210^{\circ}$ and $330^{\circ}$ in azimuth angle $\phi$ near $77^{\circ}$ in polar angle $\theta$. This effect can be observed in the simulated diffractograms in Figure~\ref{Figure XPD measurement}~(a-b). Te atoms, located in the lower plane of the Te-Ir-Te sandwich immediately above the emitter in Figure~\ref{Figure Structure}, move towards the normal emission in the $(5\times1)$ phase during the phase transition and cause this red signature in the diffractogram in Figure~\ref{Figure Exp Azimuth}. Finally, the lower central part of the diffractogram remains white indicating that minimal changes occurred in this region, as already pointed out earlier in the discussion of Figure~\ref{Figure XPD measurement}~(b). Figure~\ref{Figure Exp Azimuth}~(c) shows a diffractogram composed of the experimental XPD diffractogram for a dimer emitter in the $(5\times1)$ phase at $250$~K minus the experimental XPD diffractogram for a monomer emitter in the $(1\times1)$ phase with azimuthal cuts (d) similarly to (a) and (b). A pronounced shift in intensity of a few degrees close to $150^{\circ}$ in azimuth near $52^{\circ}$ in polar angle $\theta$ occurs, as observed before. We note that the differences between XPD diffractogram for a monomer emitter in the $(1\times1)$ phase and XPD diffractogram for a dimer emitter in the $(5\times1)$ phase are globally larger (with greater amplitudes) than the diffractogram in Figure~\ref{Figure Exp Azimuth}~(a) due to the huge displacement of the dimerized Ir atoms emitter with respect to the RT structure. An unaffected domain, in white,\CMs{lays}\CM{lies} in the center of the lower part of the XPD diffractogram difference, but considerably smaller than that in Figure~\ref{Figure Exp Azimuth}~(a), supporting the greater amount of changes in the local environment of the dimer emitter. This effect can be described as an apparent local disordered environment of dimerised Ir atoms. In addition to these difference plots, animations, presenting the XPD diffractograms one after the other, are available in supplementary materials to this paper. These animations highlight the differences between the XPD diffractograms of the different emitters in a very visible way.


\subsection{Discussion}

In our study we can confirm, within the limit of the surface local environment sensitivity of our measurement, the bulk-sensitive observations made with X-ray diffraction (XRD) for the $(1\times1)$ phase~\cite{JobicZeit,KoNatCom} and the $(5\times1)$ phase~\cite{PascutPRB,TakuboXRD,KoNatCom} as well as the surface-sensitive observations made with STM~\cite{MauererPRB,HsuPRL,KoNatCom,LiNatCom,DaiPRB}. We further contribute by probing the local real space of Ir atoms down to three layers, giving access to the surface as well as to the subsurface region. We observe the break of the 3 fold symmetry induced by the phase transition, using the crossing of the Kikuchi bands as reference. We investigated this effect on both experimental XPD diffractograms for a monomer emitter and dimer emitter in the $(5\times1)$ phase. Qualitatively our results are consistent with the atomic positions obtained from the bulk structure of the both $(1\times1)$ and $(5\times1)$ phases and therefore the stacking in a staircase scheme~\cite{JobicZeit,PascutPRB} of the dimerised Ir atoms near the surface, see Figure~\ref{Figure Structure}. In particular, the XPD patterns evolution through the phase transitions confirm the displacements of the Ir atoms. This displacement leads to an apparent local disordered environment for the dimerised Ir atoms, resulting in XPD diffractograms with broad intensities. In addition, our XPD measurements  confirm that Te atoms also undergo large displacements relative to the Ir emitter, suggesting their participation in the stabilisation of the Ir dimers, as also established using DFT calculations~\cite{SalehEntropy,LiuPRB2020} and experimental investigations~\cite{OhPRL,LiuPRB2020}. The simulated diffractograms permit to estimate these displacements with a accuracy of about 5\% of the lattice parameters. Beyond this limit, the simulated diffractograms no longer display any clear similarity with the experimental diffractograms.
\\

Given the complexity of the IrTe$_2$ structure at low temperature, with multiple inequivalent emitters arranged in staircase-like scheme, XPD calculations are challenging to achieve using codes such as Multiple Scattering package for Spectroscopies using electrons to probe materials (MsSpec) code~\cite{S_billeau_2006} or electron diffraction in atomic clusters (EDAC) code~\cite{Abajo_PRB_2001}. Therefore, we leave this effort for future studies.
\\

Furthermore, we were able to measure a crystal of IrTe$_2$ at low temperature in a preferential orientation, although the photon beam size and the method of XPD acquisition implies that we probed likely a region of several hundred micrometres, much larger \CM{than} domains observed in the literature~\cite{MauererPRB,HsuPRL,ChenPhysRev}. This very large region of the $(5\times1)$ phase in a preferential orientation suggests that our crystal is locally under strain possibly due to the silver epoxy glue used, as was already observed for the $(6\times1)$ phase on a strain induced sample holder~\cite{NicholsonComMater2021}.

\section{Conclusion}

In this work, we have observed in real space the break of symmetry undergone by IrTe$_2$ through the first low-temperature phase transition using the X-ray\CMs{photoemission}\CM{photoelectron} diffraction technique. The system switches from a trigonal structure in the $(1\times1)$ phase to a monoclinic structure in the $(5\times1)$ phase resulting in large atomic displacements. Using simulated diffractograms, we identify particular changes in the structure of IrTe$_2$ observed in experimental XPD diffractograms. Our work thereby establishes a solid basis for time-resolved XPD studies to probe the ultrafast dynamics of the surface atomic structure. Such information is particularly relevant for comparison with time-resolved ARPES experiments. Using a specific photoexcitation, we expect to photoinduce at low temperature the phase transition from the $(5\times1)$ to the $(1\times1)$ reconstruction~\cite{MonneyPRB,IdetaSciAdv}. This would be for instance expressed, according to our results, in the realignment of the Kikuchi band crossing due to the recovery of the 3 fold symmetry.
\\

\section*{Acknowledgements}

This project was supported from the Swiss National Science Foundation (SNSF) Grant No. P00P2\_170597. A.P. acknowledges the Osk. Huttunen Foundation for financial support.

We acknowledge the Paul Scherrer Institute, Villigen, Switzerland for provision of synchrotron radiation beamtime at beamline PEARL of the Swiss Light Source. M.R. and C.M. would like to particularly thank P. Aebi and T. Jaouen for the fruitful discussions. Skillful technical assistance was provided by F.~Bourqui, B.~Hediger, J.-L.~Andrey and M.~Andrey.

{\small{\textbf{\textit{This is the version of the article before peer review or editing, as submitted by M. Rumo to J. Phys.: Condens. Matter. IOP Publishing Ltd is not responsible for any errors or omissions in this version of the manuscript or any version derived from it. The Version of Record is available online at \href{https://doi.org/10.1088/1361-648X/ac3a45}{DOI/10.1088/1361-648X/ac3a45}.}}}}


\section*{Bibliography}
%

\end{document}